# Personalized Web Services for Web Information Extraction

Z. Jarir, M. Quafafou and M. Erradi

**Abstract—The field of information extraction from the Web emerged with the growth of the Web and the multiplication of online data sources. This paper is an analysis of information extraction methods. It presents a service oriented approach for web information extraction considering both web data management and extraction services. Then we propose an SOA based architecture to enhance flexibility and on-the-fly modification of web extraction services. An implementation of the proposed architecture is proposed on the middleware level of Java Enterprise Edition (JEE) servers.**

## I. Introduction

THE Web is now considered as the world's hugest database. However the web data, generally in the form of HTML pages, is destinated to be viewed in a browser by human users. The languages in which data are given are presentation languages which give no idea of the semantics of the data contained in these pages. Therefore this enormous amount of data is seemingly useless. Thus, the field of information extraction from the Web emerged to increase the automatic access to such information. We need to increase web automation by allowing applications to access directly to such huge Web information. Existing Web Services might be able to respond to such information needs. However existing web services are black boxes which have been hard-coded in the sense that they have been written once and for all in a given language. In order to be able to respond to a wide variety of information needs one needs flexibility. Building a specific service responding to a specific need should be made easy. This can be done by decomposing an information extraction task into smaller simple tasks which can easily be automated. Thereafter and according to users' requirements, a new added value Web for information extraction can be recomposed and/or reconfigured to build the required information extraction service. Existing research on Web Services has considered the problem of composing Web Services together [1]. In the case of information extraction tasks, however, there is a need to be able to compose with specific components which are Services oriented, for example wrappers for accessing data sources which do not provide Web Services.

In this work, we have chosen Service-Oriented Architecture (SOA) to experiment our idea to build personalized composite services. This is because SOA is an emerging architectural style that provides the basis of the next generation of distributed software systems. This architecture has the advantage to address problems related to the integration of heterogeneous applications. Also it defines systems that allow the linking resources (i.e. independent service providers) on demand using an XML-based instantiation. In our work, we have chosen JEE framework, which is nowadays one of the most well-known connection technology for implementing SOA.

However one of the strengthens of SOA is its capacity to adapt the new added value Web to offer an ad-hoc composite Web services according to the user's requirements. This capability to accommodate this composition of the needed services to the user's preferences such as technological restrictions, mobility requirements, resources constraints, etc. is called personalization [2,3,4,5]. The aim of Web personalization consist in offering the possibility to adapt dynamically or statically services to build a required Web applications related to the user's needs.

To deal with different personalization concerns that must be seamlessly integrated and in order to assist users to benefit from tailored Web content, our approach is based on middleware layer to identify and manage the diversity of personalization concerns. In our context, the middleware chosen is Enterprise JavaBeans (EJB) of Java Enterprise Edition (JEE) platform. On the other hand, and to introduce runtime adaptability in middleware environment, we opt for the computational reflection [6,7], which is a suitable technique for dynamic adaptation features.

The contribution of this paper is fourfold: (1) description of the web information extraction services and their orchestration, (2) proposition of architecture for services orchestration at the middleware level (JEE servers), (3) the personalization is done by searching for Web services having the required input and output parameters related to the user's request. (4) If the exact match does not found, our approach consists also on composing dynamically Web services to handle user's query.

## II. Web Information Extraction

A typical information extraction system supports querying a source, retrieving the result pages and extracting the results from them by applying a program transforming the result pages into a machine readable format.

Zahi Jarir is now with Department of Computer Science, Faculty of Sciences Semlalia, Cadi Ayyad University, Marrakech, BP. 2390 Avenue Prince My Abdelah Marrakech Morocco (e-mail: zahijarir@ucam.ac.ma).

Mohamed Quafafou is with the Information and System Science Laboratory (UMR CNRS 6168), Université de la Méditerranée, ESIL, Luminy case 925 - 13288 Marseille cedex 09 – France (phone: +33 491-828-673 , fax: +33 491-828-671; e-mail: mohamed.quafafou@univmed.fr).

Mohammed Erradi is now with Computer Networks and Multimedia Research Group, UFR-RT, ENSIAS, Mohamed V Souissi University, B.P. 713, Agdal, Rabat, Morocco (e-mail: erradi@ensias.ma).

*A. Generic Services for Web Data Management*

Some of information extraction services are generic as they are associated to web information management including data source querying, web pages fetching and parsing. This class of generic includes the following generic services:

*1) HTTP query building:* An HTTP query is composed of three parts: a query method, a base URL and a set of key/value pairs. This service builds these three parts from its parameters and returns a list containing a unique item: the HTTP query.

*2) Fetching:* A fetching service takes as input either a URL or an HTTP request and proceeds to the downloading of the document referred to. It returns an HTTP response or an empty list in case of an error.

*3) Querying:* A service querying consists calling a predetermined service with a set of parameters. It takes as input a set of parameters and outputs the result of calling the service. A service can be implemented as a web application or a web services.

*4) Parsing:* A parsing service takes an document with a specific format like XML, HTML, PDF, DOC, etc., parses it and returns a result according to a specific type like DOMobject, abstract type, etc or an empty list in case of a parsing error.

*5) Filtering:* This service does a selection on its input according to a predetermined predicate that can be defined as a set of tests. Any input object verifying the predicate is returned. All other input is kept back.

*6) Extracting:* An extraction service returns subparts of its input using an expression which is applied to the input. For example, given the DOM representation of an HTML page and the //a/@href XPath expression, the resulting extraction service returns the links contained in the input document.

*7) Transforming:* A transformation service consists changing the format of the input. When the input is an HTML/XML document (or its DOM representation) the transformation can be described by an XSL Stylesheet.

*B. Web Information Extraction Services*

We have previously presented an extracting service based on expressions which identifies the appropriate objects to extract. Unfortunately, the extraction service may be more complex as the research field on web information extraction is active and many algorithms for wrappers construction are available. We classify these algorithms into the three following main classes:

*1) Pages labeling:* The input of the algorithms of this class is a labeled page example where each and every instance found on these pages needs to be labeled. Kushmerick has introduced and formalized in [8]. Therefore the method is incapable of learning a wrapper for many sources. In order to improve expressiveness a method named SoftMealy which learns a transducer from labeled pages [9]. Muslea et al. consider the problem of hierarchically formatted pages [10] and propose the Stalker system. In this case an extraction rule needs to be learned for each level in the hierarchy. Finally, [11] propose to detect page format changes and a method to label pages in order use Stalker to learn a new wrapper. The main limit here is that the labeling process is a manual one.

*2) Document structure analysis:* Some authors consider that a wrapper can be obtained by the analysis of the “logical” structure of the pages. In [12], authors propose to automatically construct a wrapper by using the search for maximal prefixes of the sequence formed by a document. In [13] is proposed an approach to learning wrappers capable of extracting lists from tables by searching for a template among a set of pages (common subsequences of the documents). Finally, the system Roadrunner [14] uses a similar approach which consists in taking one page and trying to match the other pages against the current wrapper.

*3) Knowledge-based wrappers:* Other research [15, 16] involves the construction of knowledge based wrappers. Rather than allowing extracting data from specific sources, knowledge-based wrappers have as objective to extract data from any source of a given domain. The domain knowledge these methods rely on is however often very related to the sources from which to extract the data and therefore need a thorough analysis of different sources of the domain.

*C. Web Information Extraction Task*

In order to build a complete information extraction task it is necessary to coordinate the basic tasks. This is simply done by telling each task what to do with its results. For example, after having built a query, the next step is to fetch the query result. This can be done by setting up a query task and a fetching task and telling the query task to send its results to the fetching task. Whenever the query task receives input and builds a new query, it then sends the generated query to the fetching task.

For instance, consider the use of Google via its web services. The objective of this task is to obtain the modification date, size and type of the results given by Google to a query. The results are obtained by using Google’s doGoogleSearch Web Service. However they do not contain the information wanted which is the type of document, the last modification date, and the content size. This information can be obtained by querying the server on which the page can be found by sending an http HEAD request. To resolve this task, we first need a Web Service querying operator which knows where the Google service is located, which method to call and how to translate incoming data into a suitable parameter list for the web service call. Secondly, we need a XML parsing operator to give us a DOM representation of the obtained SOAP message. Then we need an extraction operator knowing how to extract from this message the list of result URLs. To obtain the information on each of the URLs, a fetching service is necessary to query the host server of the document pointed to by each URL. Finally, we need an extraction operator to keep for each result the desired information (ie. the url, its modification date, its size and its type). Figure 1 gives the coordination graph of this task.

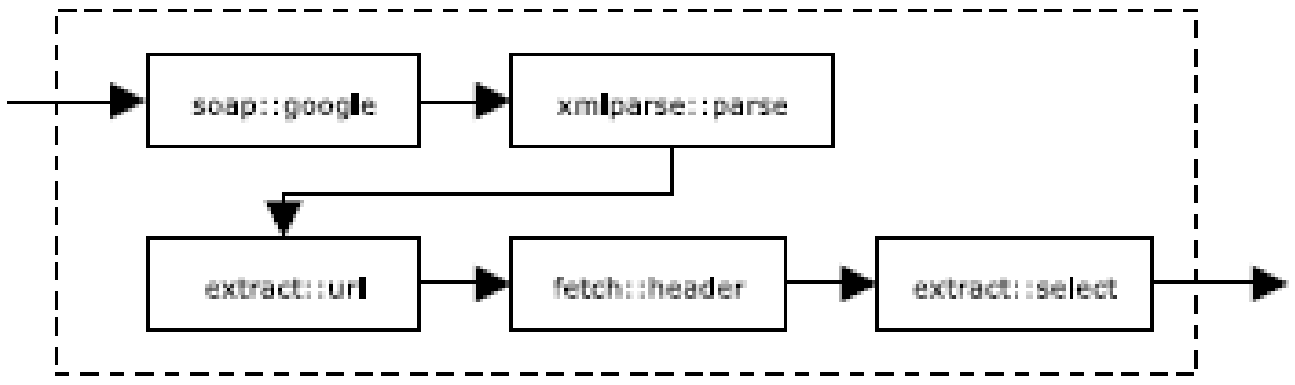

Fig. 1. Google extraction task

First of all, we need to describe the set of operators. In our XML language each operator is associated to an XML element. Each operator can be setup by declaring the values of a set of parameters by adding child elements to the operator element. The name attribute of the element gives a name of the operator. The resulted language WetDL (Web extraction task Description Language) allows describing instance of the generic services introduced previously and coordinating them building a network of service [17]. Thus to each service corresponds an XML tag name. For example the query tag name declares a query building service. The service descriptions have some common proprieties which are described by a common set of XML attributes in the operator tags.

### D. *Wrappers Learning*

In the recent past years many researchers have tackled the problem of building wrappers for such sources. The state of the art approach is to use machine learning approaches based on fully labeled example pages. In this paper we propose and study an approach based on example instances. This allows the user to build a wrapper using only a handful of examples of the whole source allowing taking into account structural differences. The patterns obtained allow extracting the instances of the relation described by the examples and contained in the same data source. The method presented uses pattern generalization, initially based on the contexts of the given examples.

The main advantages of our method are that it is efficient, incremental, generic, and robust faced to the diversity of web sources.

### A. *Context Generalization*

The approach we take is based on the generalization of contexts. We consider the user wishes to extract a given relation from a given source. To describe this relation the user only needs to give a small set of example instances. Having instances as examples relieves the user from having to fully tag example pages while still precisely defining the desired relation.

Building the extraction patterns allowing extracting a given relation for a source can be done in a three phase process: (1) a preprocessing phase, (2) a context extraction phase and (3) a pattern generalization phase. In the preprocessing step the selected documents of the source are cleaned and encoded. This allows having a more uniform format to work on in the discovery phase. In the context extraction phase the user given example instance a searched for in the documents of the target source. Their contexts are then extracted as the first patterns of the pattern generalization phase. The set of the resulting patterns is initialized by the contexts of the given example instances extracted. Once the result set is initialized a mining step can be entered which consists in generalizing these contexts into more general patterns representing the relation the user wishes to extract.

### B. *Experimental Results*

We have implemented our method in a system called IERel (Information Extraction of Relations) and applied it to different web sources of different domains such as ecommerce sites, online address books, search engines, etc.

Some of the results we have obtained for these sources are given in figure 2. The columns respectively contain the source names (Source), the number of given examples (Ex.), then number of instances considered (Inst.) the number of instances retrieved with the generated patterns (Retr.), the calculated recall (Rec.) and the calculated accuracy of the generated wrapper (Acc.). In all the cases the accuracy is of 1.00 meaning that the method did not extract erroneous instances. This is due to the fact that our generalization methods prefer having multiple specific patterns than one too general pattern.

This is obtained by being very strict (match or fail) on the generalization of tags which give string structural clues. Considering the recall, not all results are equal to 1.00. These often usually mean that the given examples were not sufficient to learn a wrapper extracting all the data. As we will see this is due to the multiplicity of formats in which a source may present its results. The regularity of most sources allows considering that it is very often possible to learn a perfect wrapper for a given source. Therefore an interesting aspect of example-based wrapper construction is the number of examples required in order to build a perfect wrapper. We claim that a wrapping a web source gets more difficult when the data to be extracted is presented in many formats.

| *Source* | *Ex.* | *Inst.* | *Retr.* | *Rec.* | *Acc.* |
|---|---|---|---|---|---|
| BigBook | 5 (1) | 4299 (235) | 4299 | 1.00 | 1.00 |
| OKRA | 5 (1) | 3780 (252) | 3780 | 1.00 | 1.00 |
| AddressFinder | 10 (1) | 88 (10) | 88 | 1.00 | 1.00 |
| AltaVista | 3 | 50 | 50 | 1.00 | 1.00 |
| Excite | 4 | 10 | 9 | 0.90 | 1.00 |
| Galaxy | 2 | 20 | 20 | 1.00 | 1.00 |
| Overture | 2 | 15 | 13 | 0.86 | 1.00 |
| Metacrawler | 4 | 30 | 28 | 0.93 | 1.00 |
| Savvysearch | 5 | 12 | 11 | 0.92 | 1.00 |
| Webcrawler | 4 | 20 | 19 | 0.95 | 1.00 |
| Google | 3 | 100 | 72 | 0.72 | 1.00 |
| Alapage | 3 | 13 | 13 | 1.00 | 1.00 |
| Conforama | 3 | 6 | 6 | 1.00 | 1.00 |
| Darty.fr | 4 | 10 | 10 | 1.00 | 1.00 |
| Fnac | 5 | 13 | 13 | 1.00 | 1.00 |
| Ikea.fr | e 3 | 15 | 15 | 1.00 | 1.00 |

Fig. 2. Extraction results on different web sources

## III. PERSONNALIZATION DIMENSIONS

WEB services operate in a cut-throat environment where even satisfied users and growth do not guarantee continued existence. As users become ever more proficient in their use of the web and are exposed to a wider range of experiences, they may well become more demanding, and their definition of what constitutes good service may be refined. Personalization is an ever-growing feature of on-line services that is manifested in different ways and contexts, harnessing a series of developing technologies [3].

The main goal of Web personalization is to facilitate the expression of the need of a particular user and to enable him to obtain relevant information when he accesses an information system. The relevance of the information is defined by a set of criteria and preferences specific to each user or community of users. These preferences may concern various aspects such as technological restrictions, mobility requirements, resources constraints, etc. The data describing the users' interests and preferences is often gathered in the form of profiles.

The content of a profile varies according to the target application (databases, information retrieval, digital libraries, multimedia applications, telecommunication services, etc.) and to its context such as fluctuating network bandwidth, and/or associating the Web application to a specific terminal, and so. For instance, certain users may wish to have access using their PDA to a YouTube service to extract, browse and either personalize information published by a website. In addition they may prefer that the required web service will be adapted on-the-fly according to continuous variations of the execution environment including contexts related to theirs configurations and also to the web services. For example if their PDA performance are decreasing, the composite web service will be adapted dynamically by eliminating video web service from the composition.

A personalized service need not be based on *individual* user behavior or user input. The content of a website can be tailored for a predefined audience, based on previous research of a defined community, and providing different sections on the website for each audience identified. This approach would give tailored content without explicitly building the one-to-one relationship that requires gathering knowledge on individuals.

### A. *Service Customization features*

To insure several requirements related to interoperability, portability and reusability of software components and separation of service from specific technologies, and managing a complex system among different business stakeholders, such as consumers, service providers, and connectivity providers, TINA architecture is based on four principles [18]. These principles are Object-oriented analysis and design, Distribution of service software components, Decoupling of Software components and Separation of concern.

An important study was done by TINA (Telecommunication Information Networking Architecture) consortium that consists to classify customization of services as shown in figure 3.

TINA service customization considers three different kinds of actor (End-User, Subscriber and Service Provider) that are responsible for performing such customization [19]. It considers also three essential aspects which are:

- Service setting that concerns the specific features and their possible values to be offered to users.
- Usage constraints under which a service can be executed
- Configuration requirements in terms of terminal and network access to support the service.

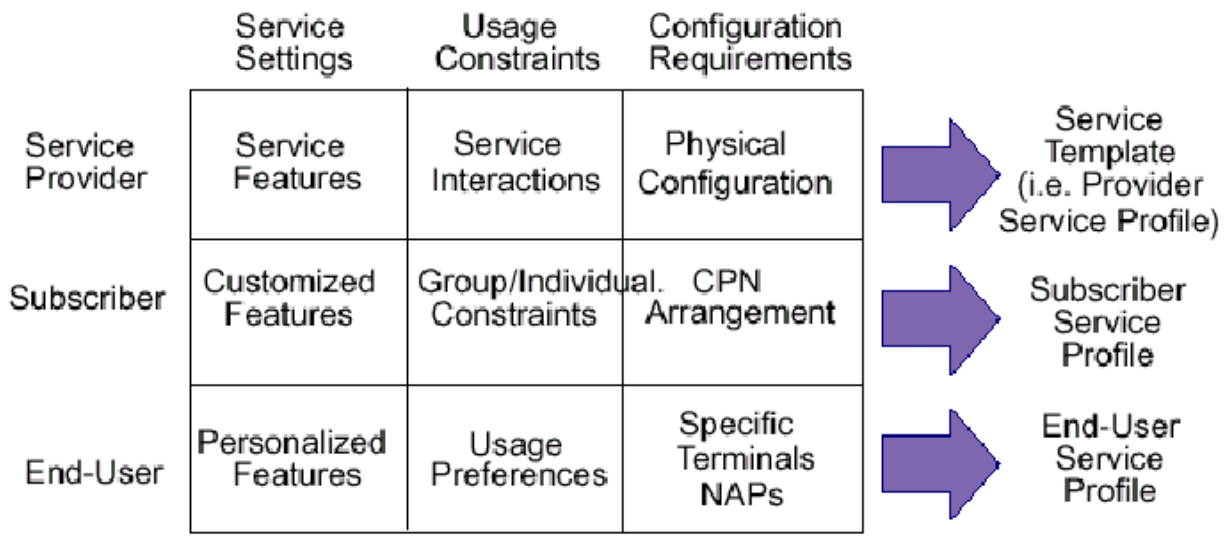

Fig. 3. Customization of TINA services

The Service Provider may customize the characteristics of a service by modifying the service template. The modification of the service template affects the service which will be instantiated after such modification. The Subscriber may customize the characteristics of a service by modifying the subscriber service profile. The modification of the subscriber service profile affects the service which will be instantiated after the modification. There are two ways to customize the characteristics of a service by the end-user: The first one consists in modifying the end-user service profile which is similar to the subscriber customization. The second one deals with the modification of the customizable data which has bundled to a service instance.

For example, a Service Provider can refine a service template by adding a set of service aspects or features. Moreover certain constraints on the execution of the service can hold (they are related to service interactions: a certain service cannot be executed in a combination with other services, or certain features are exclusive). A physical configuration is required in order to provide a service (e.g. some terminal or a specific network).

### B. *Web Personalization Taxonomy*

Personalizing Web services is therefore a challenging task and becomes even more powerful when applied to advanced and complex Web services. In this area, a new paradigm of Web engineering, called engineering adaptive Web, have been emerged to take into account the special features and needs for the development of adaptive Web applications and services [20, 21].

One of the aims of this paradigm is to propose new methodologies, tools and architectures for personalization, especially dynamic personalization, to build Web applications and services that can be automatically adjusted to varying client environment and user preferences. Several approaches in this direction are currently investigated, using different techniques and methodologies, in different areas e.g. eCommerce [22,23], Distributed eLearning Environments [24,25], Telecommunication Services [19,26], etc. Basically the major contributions of these approaches are directed mainly towards the three principal following topics: information presentation and service.

*1) Web Information Personalization:* It focuses on searching and browsing the Web information using personalized Web search systems. Recently researchers are interesting to define how to enable personalization functionality to personalize the interaction with web content [25,4,5]. The adaptive Personal Information Environment (a-PIE) based on the Semantic Web is an example. The aim of this approach is to provide the adaptive information for particular users' needs, to reuse and share information for an individual. a-PIE provides an information environment for users in a community or organization in which to browse or search information based on domain concepts defined by an ontology. The users are also able to manipulate their own information space by adding or deleting data or parts of information structures into their own information space. In addition, they can add personal information such as comments or notes to the existing data or information structures while the original data or information structures published remain unchanged [27].

*2) Web Presentation Personalization:* It contributes to tailor the information presented and the structure to the user's preferences, knowledge or interests. This topic focuses on how adaptation can be implemented through the manipulation of links, and content and presentation of nodes [28,23,29]. Among W3C works in this category, we cite Personalized Information Description Language (PIDL) [30]. The purpose of PIDL is to facilitate personalization of online information by providing enhanced interoperability between personalization applications. PIDL provides a common framework for applications to progressively process original contents and append personalized versions in a compact format. PIDL supports the personalization of different media (e.g. plain text, structured text, graphics, etc), multiple personalization methods (such as filtering, sorting, replacing, etc) and different delivery methods (for example SMTP, HTTP, P-multicasting, etc).

In addition, one sub-category of this context is terminal adaptation that assumes to provide different user interface addressing heterogeneous capabilities of device classes [31,32].

*3) Web Service Personalization:* that contributes to tailor the Web services structure and/or behavior to satisfy specific user requirements. To address this kind of personalization there is mainly two ways:

- *Using conceptual approaches* that consist in modeling and designing personalized Web applications. The aim of this approach is to bring dynamic personalization support in Web conceptual modeling constructs by creating new conceptual modeling approach or extending the existing ones e.g. OOH, OOHDM, etc. [33].
- *Using implementation approaches* that consist in modifying the implementation details of the developed Web applications. These approaches introduce personalization features using different techniques such as mobile agent, reflection techniques, and so on [34].

### C. Discussion

Most of the evoked researches focus mainly to a specific Web application concern such as building tailored presentation interface, showing personalized information, adapting application's functionalities, etc. However less attention has been paid to build architecture that covers almost all the concerns of Web personalization.

In this paper we present architecture for personalizing Web services. This architecture has the advantage to take into account at the same time the Web presentation personalization, Web information personalization and the Web service personalization based on the implementation details rather than using conceptual approaches. Our argument is that personalization requirements change over time and thereafter it's very hard to determine at the conceptual time the required personalization concerns, especially if we take the evolvable nature of the Web applications.

Building architecture for personalizing Web application imply to deal with different personalization concerns that must be seamlessly integrated. So without the appropriate approach, the task of integrating dynamic composition of business process may become a tough task. That's why we have elaborated as a first step a deep taxonomy showing the varieties of personalization concerns. These concerns are related to the following changes:

- *Parameters of the application* that concerns the application's data, as for example QoS parameters of the multimedia applications (e.g. debit, video resolution, etc.)
- *Functional aspect of the application* that concerns the application's behavior such as:
  - *Adaptability* that affects the application's behavior without calling new components or functionalities within the application. This personalization consists in activating and/or deactivating some of their already existing functionalities.
  - *Extensibility* that corresponds to introduce new additional behavior or functionalities in the application to answer a specific need.
- *Technological aspect* that concerns the modification of the application in order to be executed on different platforms (e.g. operating system, etc.), different types of terminals, the run-time variations of availability of certain resources such as CPU, memory, communication capacities, and so on.

Based on this taxonomy, a new problem emerges: what is the best level among the application level and middleware level to deal with the diversity of personalization concerns (diversity of users devices, evolvable execution context, etc.) ?

Recall that Web applications are usually developed over a Middleware which in turn is implemented over a protocol stack to ensure data communication and to manage the communication resources. Performing dynamic changes to Web services become a tedious task and need to consider changes at the underlying levels especially at the middleware level. The idea behind this approach is that services built on the top of the middleware rely on it for all interactions with the execution environment, so adapting the middleware allows us to indirectly adapt the Web services and applications. Moreover at the middleware level, the Web services may be adapted to a given configuration, either by replacing some components of code or by adding new components in order to enhance existing functionalities. Furthermore it is very difficult to adapt Web services to the continuous variations of their execution contexts at the application level.

In this paper, we show how, at the middleware level, Web information extraction that usually has to be updated and evolve over time, can be adapt on-the-fly to changing users requirements and preferences.

## IV. A MIDDLEWARE BASED SERVICE PERSONALIZATION

To address Web personalization on-the-fly in our approach and in order to meet user's preferences that constantly change, we have interested to EJB middleware architecture of JEE [35], which is an instance of SOA. An EJB is a specification and architecture for the development and deployment of distributed server-side, transactional, and secure business application components [36]. The EJB architecture is the basis and core of the Java Enterprise Edition (JEE), which defines the entire standardized application development architecture as well as the deployment environment. Therefore Enterprise JavaBeans servers reduce the complexity of developing middleware by providing automatic support for middleware services such as transactions, security, database connectivity, and more.

In this paper, we show the way used to extend EJB architecture to allow dynamic reconfiguration at runtime. This extension is made using an approach based on computational reflection which permits to create an architecture that can adapt itself to changes of user's requirements.

The aim of this section is to show how to introduce runtime adaptability in the EJB environment. In the next subsections, we will discuss how to make EJB adaptable, and then we describe the needed features of the computational reflection, which is a suitable technique for dynamic adaptation features. Afterwards, we present our adaptable middleware based service personalization.

### *A. Adapting EJB Middleware of JEE*

Oriented Middleware technologies such as CORBA or Java RMI have proved their suitability for standard client-server applications. However, such platforms do not provide the required levels of adaptation and/or reconfiguration that are needed to accommodate the diversity of modern distributed applications. Furthermore, the middleware is suitable to face the challenges of new types of applications, including support for multimedia, real time, mobility, etc. This motivates many middleware research groups to built new and advanced middleware technologies [37].

Sun Microsystem, OMG and Microsoft are aware of these limitations. For this reason, the current developments of the Enterprise Java Beans (EJB) [36], CORBA Components Model (CCM) [38] and COM+ [39] are proposed. The main focus of these platforms is to alleviate the application-level programmability issues by hiding, within the so-called component containers, a large part of the complexity with more declarative interfaces.

The growing popularity of EJB architecture (cf. figure 4) is due to the advantages offered to the distributed and Web-based applications, e.g. faster application development, ability to build complex applications, separation of business logic from presentation logic, application interoperability, etc. [37]. Furthermore, currently there are more than 30 implementations (free and commercial) of EJB servers and that number is increasing [37]. Therefore, we choose the EJB technology as an example of component-based middleware to prove that Web application personalization can be dealt with at the middleware layer.

However, the EJB platform does not address the needs for adaptation and extension required in several applications. Our suggested solution focuses on the EJB architecture to show how it can be extended to allow reconfigurations and extension at runtime providing an adaptable EJB infrastructure.

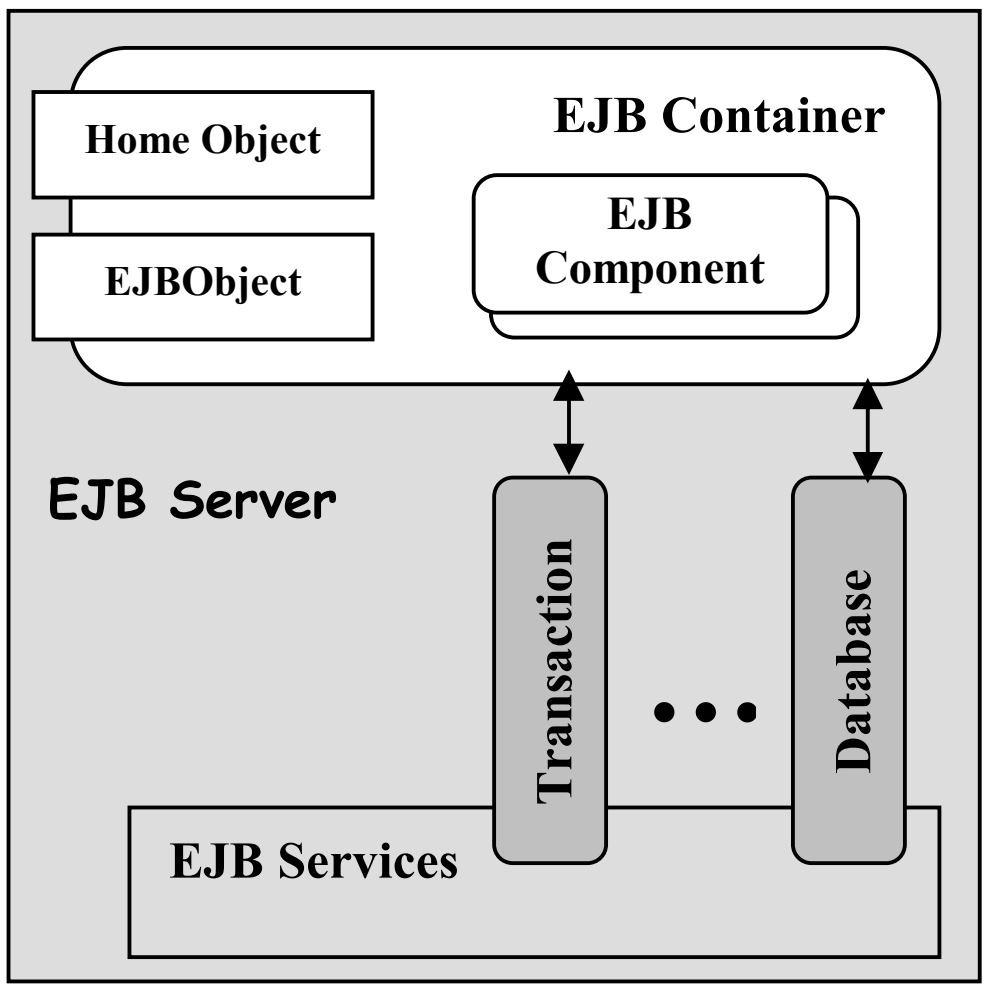

Fig. 4. EJB Architecture

According to the separation of concerns paradigm [40] used by the EJB architecture, the code of an EJB application is split in two parts: the functional code, representing EJB components, and the non-functional code, representing middleware services (e.g. transactions, persistence, security, etc.). However, the configuration between EJB components and Middleware services is only supported at deployment-time using a declarative deployment descriptor. This descriptor, presented in XML format, defines a set of accessor methods for setting and getting information about the Enterprise Java Beans being deployed.

To make EJB adaptable we need to make explicit the separation between functional code and non-functional code. Then, the associations between these two kinds of codes can be modified at runtime according to the user's requirements, to personalize its application. This reconfiguration is made by identifying which piece of the functional code is affected (and how it is affected) by the non-functional code.

To ensure an advanced adaptability of an EJB application to the desired changes, we need to focus on the EJB container layer. This is because the EJB container is an intermediary between the EJB components and the outside world especially between the EJB component and the access to various resources and EJB services. The containers are generated statically using the information provided by the EJB component's deployment descriptors. These descriptors cannot modify the associations between the EJB components and the EJB services at run-time without modifying the containers.

### *B. Computational Reflection*

Reflection refers to the capability of a system to reason about and act upon itself [6]. More specifically, it is the ability of a system to watch its computation and possibly change the way it is performed. In this direction, a reflective system is one that provides a representation of its own behavior, which can be used to inspection and adaptation and is causally connected to the underlying behavior it describes. Causally connected means that changes made to the self-representation are immediately mirrored in the underlying system's actual state and behavior and vice versa.

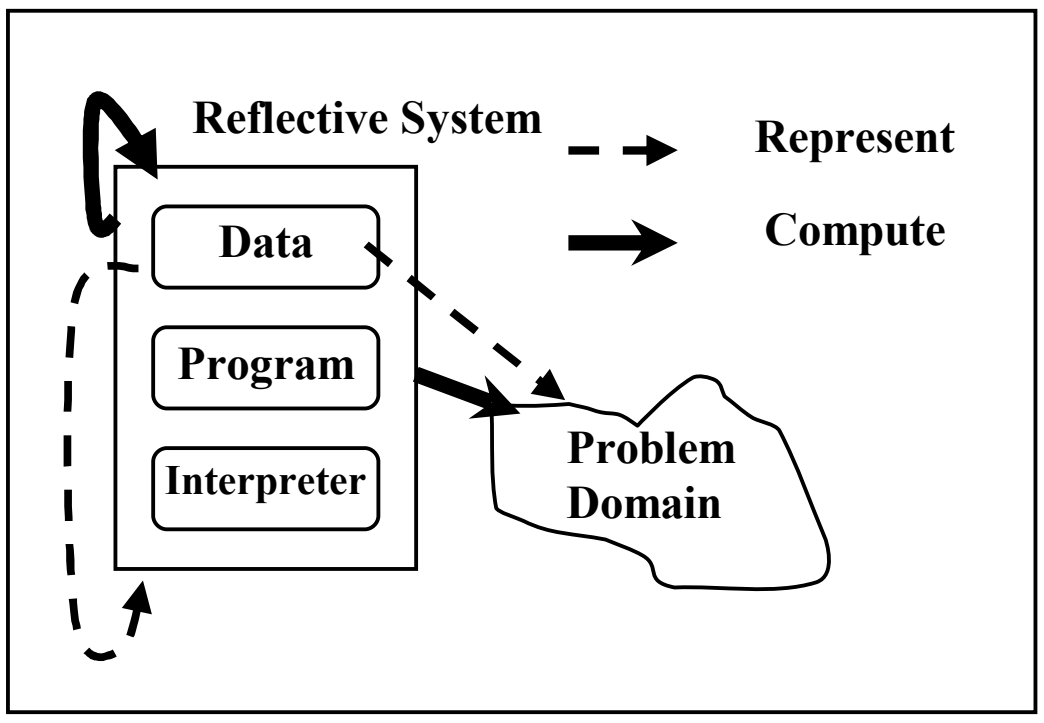

Fig. 5. Reflective system

An object oriented reflective system is logically structured in two or more levels, constituting a reflective tower (cf. Fig. 5). The first level is the base-level and describes the computations that the system is supposed to do. The second one is the meta-level and describes how to perform the previous computations. The entities (objects) working in the base-level are called base-entities, while the entities working in the other levels (meta-levels) are called meta-entities.

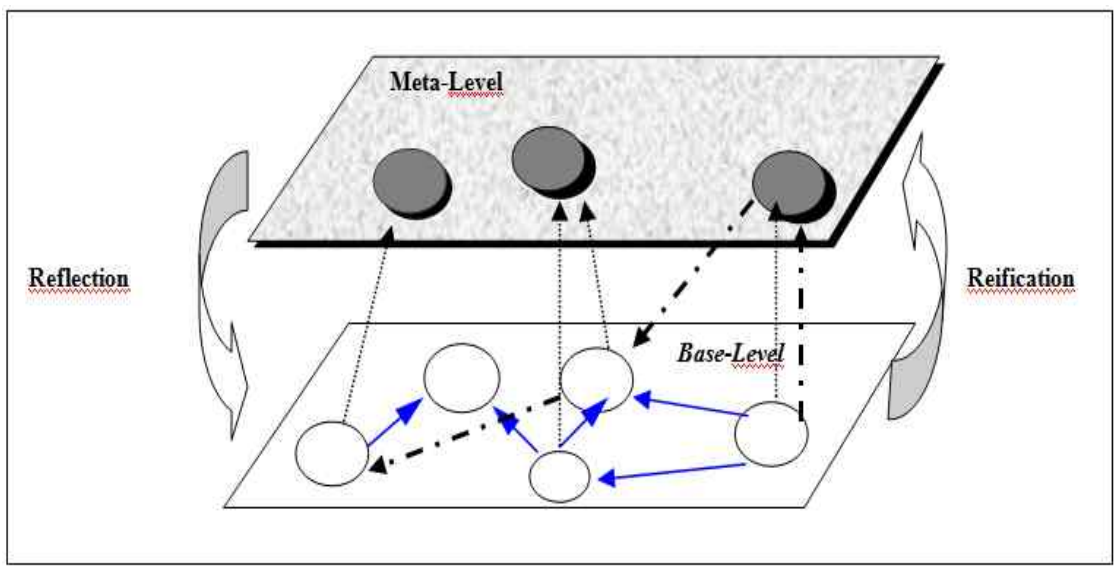

Fig. 6. Reflective Architecture

Each level is causally connected to adjacent levels, i.e., entities working into a level have data structures reifying the activities and the structures of the entities working into the underlying level and their actions are reflected into such data structures. (cf. Fig. 6).

A reflective computation can be separated into two logical aspects: computational flow context switching and meta-behavior. A computation starts with the computational flow in the base-level; when the base-entity begins an action, such action is trapped by the meta-entity and the computational flow rises at the meta-level (shift-up action). Then the meta-entity completes its meta-computation, and when it allows the base-entity to perform the action, the computational flow goes back to the base level (shift-down action).

The computational reflection has been successfully applied to several fields such as distributed systems [41], Telecommunication infrastructures [26], etc. Our experiences in this concept motivate us to choose this technique to build a reflective architecture for Web services personalization.

### *C. Adaptable EJB infrastructure*

To build our adaptable EJB middleware [42] based service personalization, we have selected JOnAS (Java Open Application Server) [43] platform. This later is an Open Source implementation of the JEE$^{TM}$ specification. It is a pure JavaTM implementation of this specification that relies on the JDK. It is part of the ObjectWeb Open Source initiative launched in collaboration with several partners including Bull, the France Telecom R&D division and INRIA. The Opening of JOnAS environment opens us the way for introducing the reflection features to try to make EJB architecture adaptable.

In addition, JonAS environment offers an open source tool, called the GenIC (Generate Interposition Classes) that allows generating the EJB container code. Therefore this tool will guarantee to set up our approach by focusing on the EJB container layer as mentioned earlier.

In order to respect the EJB container specification, we have delegated the task of a dynamic composition of services to another object called DynamicComposite, representing the meta-object of the EJB container as shown in Figure 7. The set up of this indirection is made thanks to the computational reflective features that allow diverting all methods call from EJB container to its associated DynamicComposite object. This object is able to compose dynamically attached or detached EJB services before or after sending the method call to the EJB component. Therefore the DynamicComposite object will be responsible for playing the role of a dynamic composer of the EJB services. (cf.Fig. 7).

These adaptation policies are:

- System policies, consisting in sets of rules of the form condition ⇒ action, where the condition is related to the execution environment (as reified by the monitoring framework), and the action is either the attachment or detachment of a specific services, possibly with configuration parameters.
- Application policies, which define groups of EJB components according to their runtime properties and bind existing system policies to these groups.

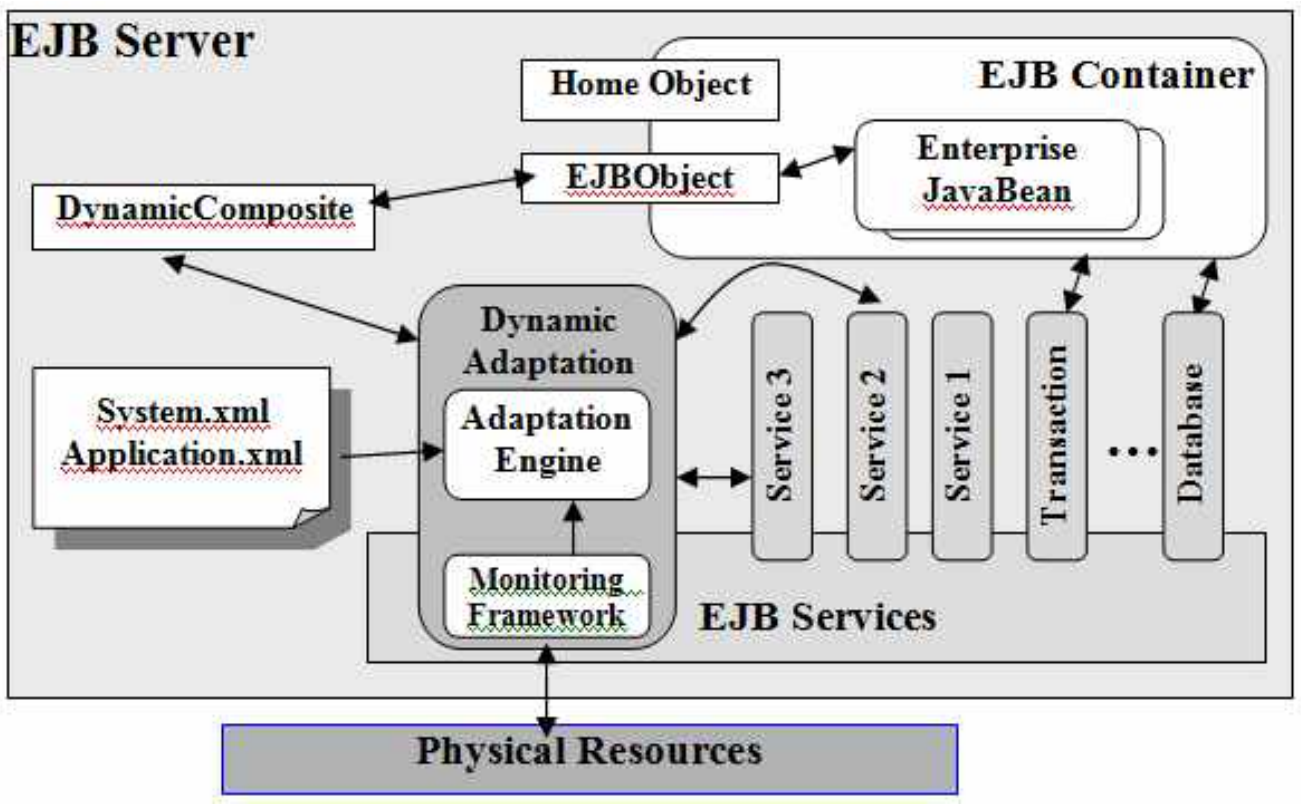

Fig. 7. An adaptable EJB infrastructure

The following XML code presents an example of a system policy named “bandwidth-policy”, which are interpreted at run-time by the adaptation engine. This policy consists of one rule that ensures to detach a service named “VideoService” from the EJB application and to update the service, named “AudioService”, by using the Sound Encoder Class “classLpc” instead of the one already used. This adaptation is performed just if and only if the network bandwidth is less than 40000 bps.

```
<system-policy name="bandwidth-policy">
 <rule>
 <when>
  <less-than>
   <property-value name="/system/network.bandwidth"/>
   <number value="40000"/>
  </less-than>
 </when>
 <ensure>
  <detached service="VideoService"/>
  <updated service=”AudioService”>
  <parameter name=”SoundEncoder”
property-value=”classLpc”/>
  </updated>
 </ensure>
 </rule>
</system-policy>
```

## V. PERSONALIZING INFORMATION EXTRACTION

According to Semantic Web paradigm in which machines can understand, process and reason about resources to provide better and more comfortable support for human in interacting with the Web, the question of personalizing the web content is at hand: Estimating the individual requirements of the user accessing the information, learning about user's needs from previous interactions, recognizing the actual access context, in order to support the user to retrieve and access the part of information from the Web which fits best to his or her current, individual needs. These needs are related in general to various criterions such as technological restrictions, mobility requirements, resources constraints, etc. These configurations may concern changes of *Parameters of the application* that concerns the application’s data, *functional aspect* that concerns the application’s behavior (*adaptability*, *extensibility*) and technological aspect that concerns the modification of the application in order to be executed on different platforms.

To illustrate our architecture, we will present a scenario showing how a user can personalize his/her information extraction (cf. Fig. 8):

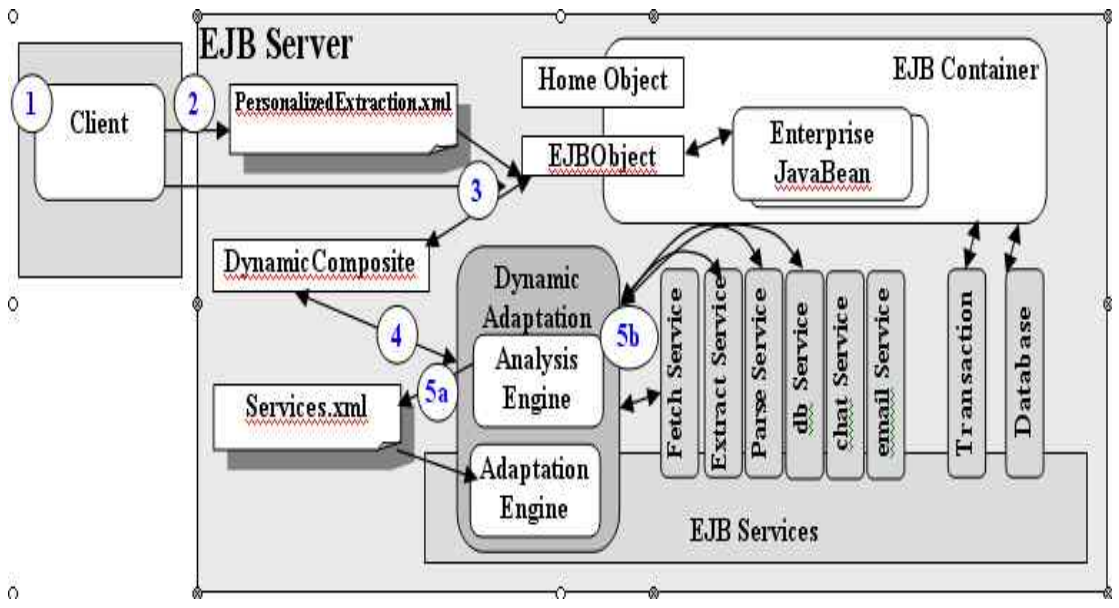

Fig. 8. General reflexive architecture

The network of tasks of the composition and the orchestration of information extraction are described in this case by WetDL language, by a file located at http://www.del-ici.fr/wsper.wdl. This network is presented as follow:

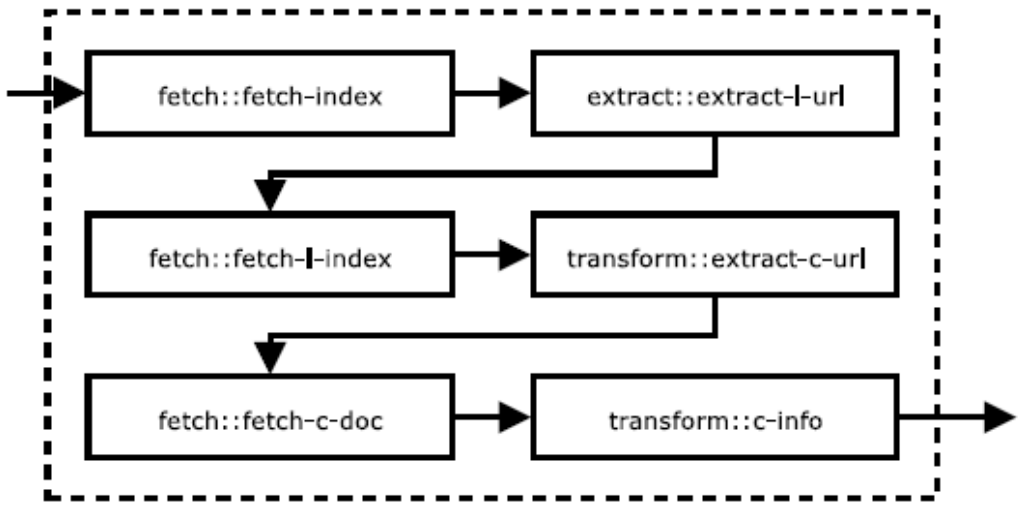

Fig. 9. Information extraction network

(2) The client program generates automatically a policy of information extraction related to the user requirement, and then will notifiy the EJBObject about this policy. This policy, called PersonalizedExtraction policy, contains the following code:

```
<PersonalizedExtraction-policy>
<updated service Sname=" dblp" Sum=" computer science bibliographie" Loc= "www.del-ici.fr" URL=" www.del-ici.fr/WetDLdblp", Slang="WetDL", Swdl=http://www.del-ici.fr/wsper.wdl />
</PersonalizedExtraction-policy>
```

(3) The EJBObject will thereafter notify its associated DynamicComposite Object about the desired personalization of information extraction

(4) The DynamicComposite object interacts with the Dynamic Adaptation service to perform the requested policy.

(5) Once the request is intercepted by DynamicAdaptation service via DynamicComposite object, the Dynamic Engine will analyses this policy to determine the required tasks that must be performed on-the-fly according to WetDL description include in "http://www.del-ici.fr/wsper.wdl". This is done by two main subsystem : Analysis Engine and Adaptation Engine. The scenario followed is:

(5a) The analyze Engine will analyze the WetDL file located at www.del-ici.fr/WetDLdblp which contains a header that defines the target web source for information extraction:

```
<?xml version="1.0" ?>
<!DOCTYPE ws:source SYSTEM
"/usr/local/share/perl/5.8.0/WebSource/websource.dtd" >
  <ws:source name="dblp.uni-trier.de">
    <ws:dummy name="init" forward-to="f1">
    <data>http://dblp.uni-trier.de</data>
  </ws:dummy>
```

In addition, it will extract, from the same file, the required services that must be attached and that compose our global web service. Also it will thereafter analyse services that are already attached according to his/her session to begin actions for information extraction. All services that are not required by services.xml will be detached. An example of this policy is:

```
<detached service name="tchat"/>
<detached service name="mail"/>
```

Where non useful service like tchat and mail are detached. Then the appropriate services are attached (parse, fetch, extract and database service):

```
<attached service name="parse">
  <parameters name="parsehtml"
        value ="instance-based"/>
</attached>
<attached service name="fetch">
  <parameters name="fetch" />
</attached>
<attached service name="extract">
  <parameters name="easy"
    value ="regexp"
    regexp="(?:[0-9]+\. )?([^/]*)(?: ?/ ?(?:[0-9]+\. )?(.*[^
    ]))? ([0-9]{4}): ([^,]+), (?:([^,]+), )?(.*)"
    map="
      <key>acronyme</key>
      <key>acronyme2</key>
      <key>year</key>
      <key>city</key>
      <key>province</key>
      <key>country</key> "
    />
</attached>
<attached service name="db">
  <parameters name="bd-dblp",
  <param name="db value="dbmasters"/>
  <param name="user" value="habegger" />
  <param name="pass" value="n62Odj" />
  />
  <query>
    INSERT INTO conference (acronyme, year, city, province,
    country) VALUES ('$acronyme',
    $year, '$city', '$province', '$country')
  </query>
</attached>
```

(5b) Once this analysis is done, the Adaptation Engine will thereafter execute the required tasks of attached and/or detached services according to services policy generated.

## VI. CONCLUDING REMARKS

In order to build a more flexible architecture for personalized Web information extraction, we have decomposed an extraction task to elementary sub-tasks which are represented by services. We have used the WetDL language to describe a composed service as a network that describes both services and their orchestration. Finally, we have proposed services based architecture to support on-the-fly modification of web extraction services without stopping the current process. We have reused the adaptable JOnAS EJB infrastructure described above to execute a required network of tasks related to the users request. Such operator is encapsulated into an EJB service. Currently, we continue the development of web information extraction services taxonomy to define the Web information EXtraction Ontology (WEXO) to support semantic based extraction services. We are also studying different solution to generalize our implementation considering not only JOnAS, but also JBOSS and .Net frameworks.

**Zahi Jarir** received his postgraduate degree in computer science in 1997 on Natural Language Processing at Faculty of Sciences in Rabat, Morocco. From 1997 to 2006, he was assistant professor at Faculty of sciences, Cadi Ayyad University in Marrakech, Morocco. In 2006, he received academic accreditation from Cadi Ayyad University. Currently, he is a professor of Computer Science at Faculty of Sciences of Cadi Ayyad University. He has participated actively to the project ARCAD to build an adaptable and reflective EJB middleware. His research interests include personalization in distributed systems, adaptive and reflective middleware, service-oriented computing, distributed programming, web and mobile applications, computational reflection.

**Mohamed Quafafou** did his PhD Thesis in 1992 on Intelligent Tutoring Systems at INSA de Lyon, France. From 1992 to 1994, he was ATER at INSA de Lyon and than at Nantes Faculty of Sciences. From 1995 to 2001, he was assistant professor at Nantes University. During that period, he developed research on Rough Set Theory, concepts approximation, data mining, web information extraction and participated actively with France Telecom to the project Comminges to design a new web system dedicated to French web analysis for discovering emergent web communities. He was also chief-scientist at GEOBS where he headed the Geobs DataAnalyzer project, which was developing a spatial data mining systems with application to environment, marketing, social analysis, etc. From September 2002, he was professor at Avignon University and moved in 2005 to the Université de la Méditerranée in Marseille where he joined the Information and System Science Laboratory (UMR CNRS 6168) and continue his research on complex web data mining and fusion. Since 2005, he teaches at ESIL foundations of data, knowledge and web engineering including machine learning, data mining, personalization, datawarehousing, XML, web services, web and mobile applications.

**Mohammed Erradi** has received Ph.D in Computer Science at University of Montreal, Canada, in 1993. He is currently a professor at ENSIAS, Mohamed V-Souissi University, Rabat, Morocco since 1993. He is head of the Alkhawarizmi Research Laboratory. His research interests include communication software engineering, cooperative mobile multimedia systems, evolving distributed systems, reflection and meta level architectures, mobile applications, telecommunication service engineering.